\documentclass[draft]{agujournal2018}

\usepackage{apacite}
\usepackage{url} 
\usepackage{comment}
\draftfalse

\journalname{JGR-Space Physics}

\begin{document}

%
%


\title{Broadband Imaging to Study the Spectral Distribution of Meteor Radio Afterglows }

%
%




\authors{S. S. Varghese\affil{1}, J. Dowell\affil{1}, K. S. Obenberger\affil{2}, G. B. Taylor\affil{1} \& J. Malins\affil{2}}


\affiliation{1}{University of New Mexico, Albuquerque, NM, USA}
\affiliation{2}{Air Force Research Laboratory, Kirtland AFB, NM, USA}




\correspondingauthor{Savin Shynu Varghese}{savin@unm.edu}




\begin{keypoints}
\item We collected broadband spectra of 86 meteor radio afterglows (MRAs).
\item The spectral index distribution of 86 events peaked at --1.73.
\item Spectral parameters are not correlated with the physical properties of MRAs.
\item The duration of the MRAs is found to correlate with other physical properties of MRAs.

\end{keypoints}

%
%

\justify
\begin{abstract}
We present observations of 86 meteor radio afterglows (MRAs) using the new broadband imager at the Long Wavelength Array Sevilleta (LWA-SV) station. The MRAs were detected using the all-sky images with a bandwidth up to 20 MHz. We fit the spectra with both a power law and a log-normal function.
When fit with a power law, the spectra  varied from flat to steep and the derived spectral index distribution from the fit peaked at --1.73. 
When fit with a log-normal function, the spectra exhibits turnovers at frequencies between 30-40 MHz, and appear to be a better functional fit to the spectra.
We compared the spectral parameters from the two fitting methods with the physical properties of MRAs. We observe a weak correlation between the log-normal turnover frequency and the altitude of MRAs. The spectral indices from the power law fit do not show any strong correlations with the physical properties of MRAs.  However, the full width half maximum (FWHM) duration of MRAs is correlated with the local time, incidence angle, luminosity and optically derived kinetic energy of parent meteoroid. Also, the average luminosity of MRAs seems to be correlated with the kinetic energy of parent meteoroid and the altitude at which they occur.

\end{abstract}

%
%

%


%
%
%
%
\section{Introduction}
Billions of meteoroid particles enters Earth's atmosphere every day with velocities from 11-72 km/s,  ablating and producing long columns of ionized plasma at altitudes between 60-130 km. The plasma trails of bright and large meteors can produce strong radio emission known as meteor radio afterglows (MRAs) at HF (3-30 MHz) and VHF (30-300 MHz) bands \citep{ob14b}. The MRAs were initially detected with the  LASI \cite<LWA All-Sky Imager;>{ob15a} correlator of the first station of the Long Wavelength Array (LWA1). 
The detected radio emission was non-thermal, unpolarized and had characteristic light curve patterns with a fast rise of 10 to 20 seconds and a slow decay which lasted up to couple of minutes \citep{ob14b}. Also, the emission was smooth and broadband between 20--60 MHz and it has not been observed below a cutoff elevation of 90 km \citep{ob15b,ob16a,ob16b}. A recent study using LWA1 and the second LWA station located on the Sevilleta National Wildlife Refuge (LWA-SV) has revealed that the emission is isotropic \citep{var19b}.

The MRAs are usually formed from meteor trails of large meteoroids ($\sim$ 0.1-10g).  Previous observations by \citet{ob15b} have shown that MRAs peak during meteor showers and some events occur from sporadic events as well. Classification of MRAs into shower or sporadic origins requires knowledge of the radiant direction. Combined radio and optical observations are required to understand the origin of meteors producing MRAs.

Currently there are two relevant hypotheses to explain the MRA emission mechanism. The first hypothesis is the possible electromagnetic conversion of the electrostatic plasma waves within the turbulent plasma trail at plasma frequencies known as the Langmuir waves. Generation of Langmuir waves likely requires a plasma instability within the trail and the possibility of this mechanism is discussed in \citet{ob15b}. The second hypothesis is the transition radiation mechanism in which hot electrons moving at a constant velocity through an inhomogenous plasma can radiate due to the difference in the refractive index of the plasma \citep{plat02}. The electrons produced during the initial ablation and ionization during the meteor entry are likely to thermalize much faster than the minute time scales of MRAs. Therefore,  both the Langmuir waves and transition radiation mechanism requires suprathermal electrons which can drive MRAs longer than the ablation timescales (few seconds).

 Using the radio data from LWA-SV and optical data from the Widefield Persistent Train Camera (WiPT),  \citet{ob20} has shown that the MRAs are temporally and spatially correlated with the long lasting (tens of minutes) emission in the optical and infrared known as persistent trains (PTs) (see, \citet{bor06} for an overview). The PTs derive their energy from the continuous exothermic chemical reactions between the ablated meteoric metal particles and atmospheric oxygen species \citep{kru01,ob20}. This work demonstrates that persistent trains can provide enough suprathermal electrons to drive the MRA emission. PTs have a long history in both the optical and radar community related to both dust and turbulence. There exist some probable connection between the PTs and the long duration non-specular echoes where the link being charged dust and turbulence. The long duration nonspecular echoes are mostly reflections from B-field aligned irregularities formed within the turbulent plasma trails \citep{ma04,dyr05,dyr08}. However, some nonspecular reflections can arise from non-field aligned irregularities as well. \citet{kel04} has suggested that the long duration radar echoes can be explained by persistent charged dust trains. Also, \citet{chau14} have shown that turbulent charged dust within the meteor trail can form non-field aligned irregularities and result in longer duration nonspecular echoes. Therefore, the presence of charged dust and turbulence in the meteor trails could be an important factor linking the formation of MRAs, PTs and long duration nonspecular echoes.  
 
 Despite various efforts to characterize the properties of MRAs, the emission mechanism is still unknown. The spectrum of a radio source provides the energy distribution as a function of frequency. The shape of the spectrum can vary depending on the emission mechanism of the source and the physical properties associated with it. Since, the emission mechanism of MRAs are poorly understood, broadband spectral measurements of the source should provide some insight into the emission mechanism. Developing theoretical models of emission mechanism requires observational constraints. These constrains can be obtained through broadband spectral measurements for a sample of MRAs. Understanding the correlation between the spectral parameters and a set of known physical properties of MRAs would help to identify the key physical properties playing a significant role in the formation of MRAs.
For the past 8 years, the all-sky imaging in both LWA stations was carried out using the LASI correlator which can produce all-sky images every 5 seconds in real time with a bandwidth of 100 kHz. The sensitivity of images is inversely proportional to the square root of the bandwidth. Due to the narrow bandwidth of LASI and limited sensitivity, the broadband measurements of MRAs in the past were carried out using the beamformed/phased array mode of LWA1. In beamforming, the digitized voltage signals from each of the 256 dipole antennas are time delayed and coherently summed to form a beam which can be pointed at any direction in the sky \citep{tay12}. Each beam can collect up to 36 MHz bandwidth of data with higher time and frequency resolution compared to the all-sky mode. This capability makes it ideal for producing dynamic spectrum of the observed sources. However, the all-sky mode has \textbf{a} large field of view of $\sim$$1 \pi \ sr$ ($\sim$$10^{4} \ \rm{deg^{2}}$) compared to $\sim$50 $\rm{deg^{2}}$ for the beamformed mode. 

In the broadband measurements of MRAs with beamforming, three beams were pointed near zenith to collect the data \citep{ob15b,ob16a}. This resulted in the detection of 2 MRAs after 5600 hours of observation in the first campaign \citep{ob15b} and 2 MRAs later in the second campaign \citep{ob16a}. The smooth spectra of radio sources are typically characterized using a power law of the form  $S_{v} \propto v^{\alpha}$, where $S_{v}$ is the flux density at frequency $\nu$ and $\alpha$ is the spectral index. The spectra of 4 MRAs from the campaigns were fitted with power law to obtain the spectral indices. The campaigns measured a spectral index of --3.8, --4.2, --4.8 and --4.4 for the four detected MRAs. 

\citet{zha18} conducted a 322 hour survey with Murchison Widefield Array (MWA) searching for intrinsic radio emission from meteors. The survey between 72-103 MHz did not identify any MRA candidates down to a 5 sigma noise threshold of 3.5 Jy/beam. However, the survey reported an upper limit of --3.7 for the spectral index (with 95\% probability of detecting at least one event) in their frequency range. This upper limit on the spectral index is higher than that of the LWA measurements. The LWA had measured the spectra of only 4 events and additional collection of spectra for a sample of MRAs are required to accurately constrain the spectral index. If the peak of the spectral index distribution for a sample of MRAs is flatter than --3.7, then MWA has a higher probability to detect MRAs in their surveys. 

One of the downsides of the beamformed observations with LWA is that the detection rate of MRAs is very low compared to the all-sky images due to the decreased sky coverage. Only four events were detected after several 1000 hours of observation which limits a statistical analysis of the broadband measurements. Also, the data calibration in both campaigns were carried out by dividing out the known instrumental responses. The lack of an astronomical calibration might have introduced some uncertainties in the spectral index measurements.

The new broadband imager at LWA-SV, known as Orville, can image the whole sky every 5 seconds with a bandwidth up to 20 MHz. The 200 times increase in bandwidth for the Orville imager compared to LASI can produce all-sky images with 14.14 ($\sqrt{200}$) times better sensitivity. Also, Orville can characterize spectral properties of the transient sources.  This provides an opportunity to make broadband measurements of MRAs with higher detection rate and introduce better constrains on the spectral characteristics.

In this paper, we present the broadband measurements of 86 MRAs from the Orville imager and a statistical analysis of the measured properties. We also utilize the optical data from Global Meteor Network (\url{https://globalmeteornetwork.org}) which are available for 28 MRAs. Section 2 describes the observations using the Orville broadband imager. Section 3 provides information about the transient search pipeline used in the detection of MRAs. Section 4 describes the calibration strategy. Section 5 and 6 presents the broadband measurement from Orville, optical data from from Global Meteor Network and the statistical analysis of MRAs. Finally Section 7 concludes the paper.

\section{Observations}
The Long Wavelength Array radio telescope located in central New Mexico currently has two stations. The first station, LWA1 ($\rm{107.63^{\circ} W, 34.07^{\circ} N} $), is co-located with the Karl G. Jansky Very Large Array (JVLA) and the second station, LWA-SV ($\rm{106.89^{\circ} W, 34.35^{\circ} N} $), is located on the Sevilleta National Wildlife refuge \citep{tay12,cra17,lm214}. Both stations have a similar physical layout comprised of 256 dual polarization dipole antennas arranged pseudo-randomly in the form of a 100 $\times$ 110 m ellipse. The LWA1 operates from 10-88 MHz and LWA-SV from 3-88 MHz frequency range. Both stations utilizes the all-sky imaging
mode by LWA All-sky Imager (LASI) or the narrowband imager which produces 100 kHz images every 5 seconds and the beamforming mode. The observations of MRAs presented in this paper were carried out using Orville, the new broadband imager at LWA-SV, which can image the sky every 5 seconds with up to 20 MHz bandwidth. The broadband imager was producing 10 MHz images every 10 seconds in the first 4 months of the commissioning phase. We have utilized 1362 hours of 10 MHz images with 10 seconds integration and 977 hours of 20 MHz images with 5 seconds integration from October 20, 2018 to April 17, 2020 for the study presented in this paper.
\subsection{The Orville Broadband Imager}
The Advanced Digital Processor (ADP) is the digital backend for LWA-SV that is based on the Bifrost pipeline framework \citep{cra17}.  In addition to supporting the same beamformed and narrowband all-sky modes available at LWA1, ADP also provides a broadband FX correlator that cross-correlates each antenna and generates visibilities every 5 s for up to 20 MHz of bandwidth. Orville is the new realtime all-sky imager for the output of the ADP broadband correlator. Orville receives the packetized visibility data from ADP, images the data and writes the images to the disk in a binary frame-based format.  The imaging is performed using $w$-stacking \citep{of14} to correct for the non-coplanarity of the array.  For each image, the sky is projected onto the two dimensional plane using orthographic sine projection. To reduce the number of $w$-planes needed during $w$-stacking, the phase center is set to a location approximately 2$^\circ$ off zenith that minimizes the spread in the $w$ coordinate. The gridding operation is based on the Romein gridder implemented as part of the EPIC project \citep{ke19,rom12}. Every 5 seconds, the imager produces 4 Stokes (I, Q, U \& V) images in 198 channels, each with 100 kHz bandwidth.  This will roughly produce 1 TB of images everyday and they are stored in the local disk. The data with reduced spectral resolution (six 3.3 MHz channels) are available at the LWA data archive (\url{https://lda10g.alliance.unm.edu/Orville/}). For more information on the implementation and data formats of Orville, see \citet{lm215}.

\begin{figure}[h]
\centering
    \includegraphics[width=\textwidth]{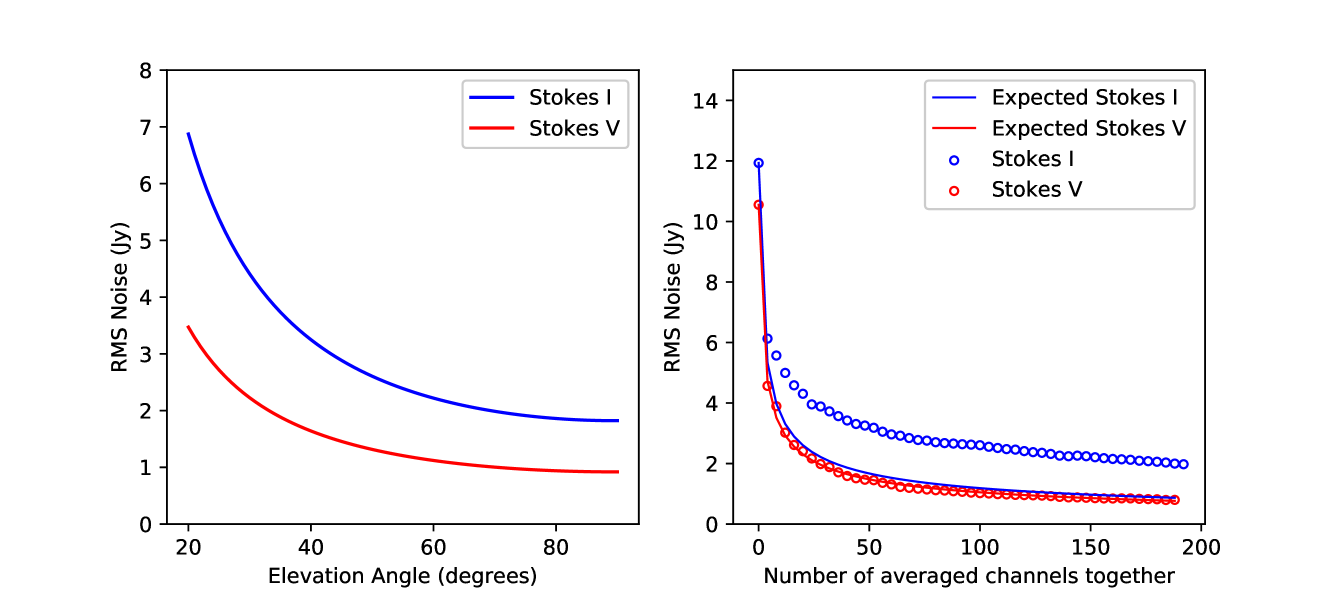} 
  \caption{ (Left) RMS noise as a function of elevation angle for the Stokes I \& V images. (Right) RMS noise of the images close to zenith as a function of number of channels averaged together for Stokes I \& V. The lines denote the expected theoretical noise and the open circles denote the observed noise as a function of number of channels averaged together for Stokes I \& V. }
  \label{noise}
\end{figure}

The left panel of Figure \ref{noise} shows the RMS noise of the images from the Orville imager as a function of elevation angle at 40 MHz. The primary beam response of a single LWA antenna has maximum sensitivity at zenith and decreases towards the horizon. The Stokes I \& V images have a maximum sensitivity near zenith and decreases at lower elevation angles. The right panel of Figure \ref{noise} shows the RMS noise of the images near zenith in Stokes I \& V and how the noise changes after averaging the images over 198 channels. The RMS noise is going down as the square-root of the bandwidth for 
Stokes V. The decrease in noise for Stokes I as a function of number of averaged channels together has some deviation from the theoretical curve. The RMS noise near zenith for the integrated image over all channels in Stokes I is $\sim$2 Jy and Stokes V is $\sim$1 Jy (1  Jy = 1 Jansky = $10^{-26}\;\rm{W\;m^{-2}\;Hz^{-1}}$). The deviation from the expected nature in Stokes I is  probably due to minimum noise level hitting the confusion limit noise around 2 Jy. The confusion limit is reached when the density of sources becomes sufficiently high enough in a synthesized beam such that they cannot be resolved. This arises due to limited angular resolution of the telescope. In Stokes V, confusion limit should be lower by at least a factor 10 as most of the sources are less than 10\% polarized. Therefore, we do not observe a deviation in Stokes V.
Also a supplemental video (broadband-movie.mp4) demonstrates the difference between a single channel image near the LASI operating frequency (38 MHz) and the combined images over all channels between 30.137 MHz to 49.837 MHz. 

\section{Transient Search Pipeline}
The transient search pipeline is based on an image subtraction algorithm. In this method, an average of the previous 4 to 6 images within the last 30 seconds is subtracted from a running image. This will give a clean subtracted image removing the contribution from steady sources and the Galactic plane.  Pixels with flux values greater than 6  times the standard deviation of the noise are marked as transient candidates. This noise threshold varies as a function of the Galactic latitude. The transient search pipeline for the narrowband imager/LASI is described in more detail in \citet{var19a}. The existing pipeline for the narrowband imager was modified to find transient sources from the new broadband images. The 198 channel images for each integration from the Orville are averaged down to a single image in four Stokes parameters. The pipeline collects these averaged images for an hour and carry out a transient search in Stokes I and V on integration time scales of 5, 15 and 60 seconds. The transient search produces roughly 500--2000 transient candidates per day in Stokes I.

Most of the transient candidate events are false positives due to the scintillation of cosmic radio sources caused by the plasma irregularities in the ionosphere. The ionosphere contains an inhomogenous  magnetized plasma and density structures which act as a screen causing refraction and scattering of the incoming radio waves from space. This  results in rapid flux changes and position shifts of the observed sources by a couple of degrees. More details on scintillation and how it affect the transient search can be found in \citet{var19a}. Two steps were carried out to reduce false positives due to scintillation of radio sources in the pipeline. In the first step, we masked all transient candidates within 3 degrees of VLA Low Frequency Sky Survey \cite<VLSS;>{Coh07, lan12} sources with flux density greater than 50 Jy at 74 MHz. The VLSS sources when scintillating have characteristic light curve patterns with rapid fluctuations and peaks over the period of half an hour to a few hours. The timescale of these scintillation depends on the ionospheric conditions at the time. For example, factors such as solar terminator, high solar activity, etc,  can induce ionospheric irregularities.  In that case signal to noise ratio (SNR) of a scintillating source from the light curve will be lower. In the second step, light curves of the events over the duration of an hour was used to filter out low SNR events which includes most of the scintillating sources.

The next step of the pipeline filters out the narrowband RFI (radio frequency interference) events such as reflections from meteors trails, narrowband transmitters, etc., using a sliding median window filter technique.  In this technique, we define a window with 11 channels (1.1 MHz) within the bandpass response of a source over 198 channels (19.8 MHz). Then moving the window across the spectrum and calculating the median within the sliding window will give a smooth bandpass model. The smooth bandpass model can be subtracted from the observed response to find deviant channels greater than 3 sigma. The bad channels are flagged from the data and events with SNR less than 5 sigma in the subtracted images and light curves were filtered out as narrow RFI events.

The final step of the pipeline removed slowly moving objects in the sky like airplanes. A single station of LWA detects several self emission and reflection events from slowly moving objects like airplanes which move from horizon to horizon. Typically an airplane at 10 km altitude smears across 10--15 degrees in a single 5 second image. Most of the MRAs observed by LWA are fast meteors with an average velocity of 50 km/s \citep{ob14b}. Even the slowest meteor with 12 km/s covers at least 34 degrees in 5 seconds at 100 km elevation. The airplane filter method uses a background image subtraction and mean shift of pixel position algorithm on the images. 
This technique is widely used in detecting the motion of objects in static cameras like traffic cameras.  An average of all-sky images 2 minutes before and after the transient detection will give a good background sky as steady sources remain almost stationary within that time duration. Also, bright sources like Cygnus A (Cyg A), Cassiopeia A (Cas A), etc. are masked during background subtraction to avoid finding steady source residuals within the search window. The background image is subtracted from the starting image and initializes the position of transient source. Then in the next image, the algorithm looks for sources greater than 5 sigma above the mean within a search radius of 20 degrees. If there is a source within the radius, the mean values for the azimuth and altitude of the new source is calculated and the search window center is moved to the new position.  This mean shift procedure is carried out within one minute time window of the event and it stops if the mean position of a source is not changing. Finally, sources which change position in more than two images are filtered out as slowly moving objects. The final list of transient candidates after filtering the low SNR, narrow RFI and slowly moving sources decreases to less than 20 events in Stokes I. Then the candidates are manually inspected to confirm their nature.

\section{Calibration of MRAs}
The MRA events were calibrated using Cyg A. Scintillation of radio sources is very intensive at low frequencies. During times of high scintillation, the ionosphere introduces frequency structures in the form of enhancements and dips in various regions of the bandpass spectrum of the calibrator source. A supplementary video shows the effect of ionosphere on the Cassiopeia A (Cas A) radio source during scintillation high (cas-scintillation-high.mp4) and quiet times (cas-scintillation-low.mp4). This variable, frequency-dependent structure results in a poor calibration of the MRA source. Therefore, in order to perform a good calibration, it was necessary to find scintillation quiet times. While most scintillation of astronomical sources is caused by structures in the F layer, it is possible that structures in the E layer (for e.g. sporadic E) could cause scintillation of MRAs. This may manifest as time-varying small scale spectral structures, similar to what was seen in \citet{ob16a}. However, correcting this effect requires a measure of E region conditions near the MRA locations which are not available for the sample presented in this paper.


An existing, widely used scintillation index \cite<S4;>{fre78}, can be used to study the ionospheric activity. The S4 index is usually measured using signal transmitting satellites and receivers on the ground. The receiver will measure the change in phase and amplitude as the signal passes through the ionosphere. \citet{mal18} used GPS satellites and receivers located at LWA-SV to study the time-varying Faraday Rotation of radio sources caused by the ionosphere. These systems can be used to calculate the scintillation index, but they operate at L band frequencies (1 to 2 GHz). The S4 index measured at high frequencies may not reflect the ionospheric scintillation experienced by radio waves below 100 MHz. Also, a measurement close to the calibrator (since the satellite orbit changes) is required to accurately model the direction dependent propagation effects through the medium. Furthermore, S4 index does not capture the frequency dependent effects as they operate at narrow frequency ranges. This motivated us to develop a scintillation index using the source responses from images which can define times at which scintillation is quiet and high.

\begin{figure}[h!]
\centering
 \includegraphics[width=30pc]{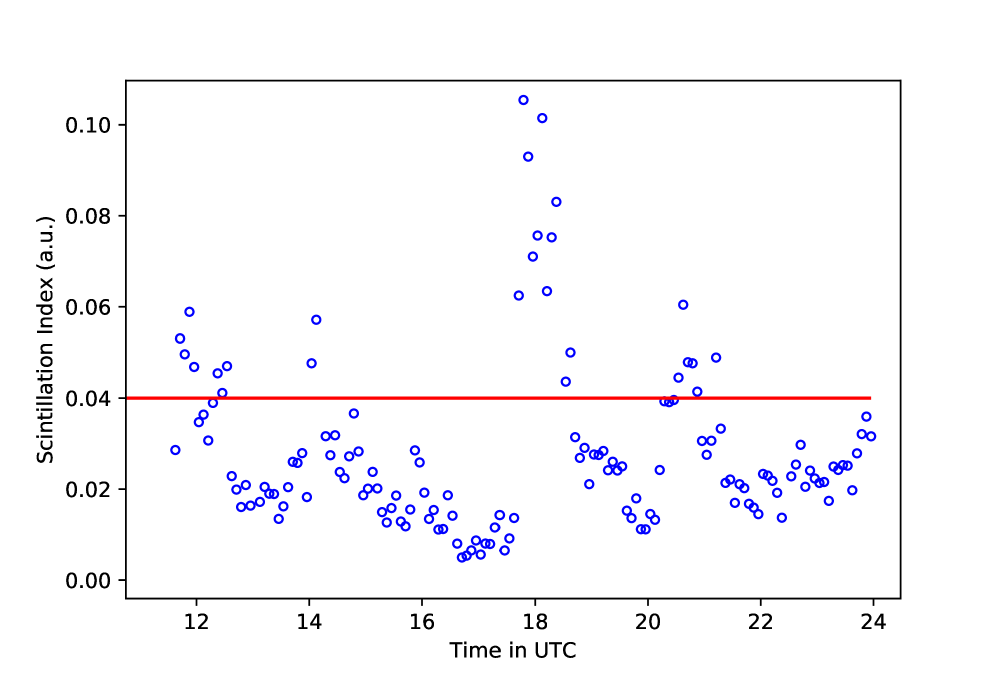}
 \caption{Scintillation index measured for Cygnus A on February 9, 2020 over several hours. Flat line denotes the scintillation index boundary of 0.04, where below the line is scintillation quiet and above the line is scintillation high.}
 \label{scint}
\end{figure}

The bandpass response of a calibrator source changes as a function of elevation as the primary beam response of a single antenna varies with elevation. 
Even though the calibrator response changes as a function of elevation, it effectively remains the same as the source moves one degree in elevation in the sky. To calculate the scintillation index, we collected the calibrator responses from a range of all-sky images as the source covers one degree in elevation (approximately fives minutes). The individual responses are divided by the median of the collected response over the particular time window/elevation. Calculating the standard deviation of the individual divided response gives an estimate of how much each response deviates from the median value. We define the mean of the standard deviation calculated from the individual divided responses to be the scintillation index for that time window. If $B_{i}$ is the measured bandpass response for each time integration within a time/elevation window and $B_{M}$
is the median value of the responses, then scintillation index can be written as
\begin{equation}
    \rm {Scintillation \; Index} = Mean \left[ SD \left(  \frac{\it B_{i}}{\it B_{M}}    \right) \right] \mbox{,}
\end{equation}
where $\rm SD$ is the standard deviation.
When the ionosphere is calm, the calibration response should be fairly constant over that time scale. Then calculating the median calibrator responses over time and dividing individual responses by the median should ideally give a flat response at all times. This will give a scintillation index of zero. After analyzing data on different days and calibrating known sources, we define the scintillation quiet times when scintillation index is less than 0.04 and scintillation high if it is greater than 0.04. Figure \ref{scint} shows the scintillation index measured on February 9, 2020 with Cygnus A showing scintillation quiet and high times. The plot shows high scintillation around 18 UTC (local noon). The flat line in the plot denotes the boundary between scintillation quiet and high times. The variation in the scintillation index is not fixed and can change on a daily basis depending on the condition of the ionosphere. This method can be also used to study the state of the ionosphere.

The average of the calibrator responses from scintillation quiet time was used for the calibration of transient sources. For the calibration of MRAs, an average of all-sky images 2 minutes before and after the peak of event was subtracted from the peak image to get a good subtracted image. Then the peak position power values were measured in each channel image to get the transient source response. This transient source response at a particular elevation was divided by the averaged calibrator response at the same elevation. The ratio of source response to calibrator response was converted to Jy units using the flux density and spectral index of Cygnus A from \citet{ba77}. 

\section{Broadband Spectra of MRAs}
\begin{figure}[h!]
\centering
\includegraphics[width=\textwidth,height=\textheight,keepaspectratio]{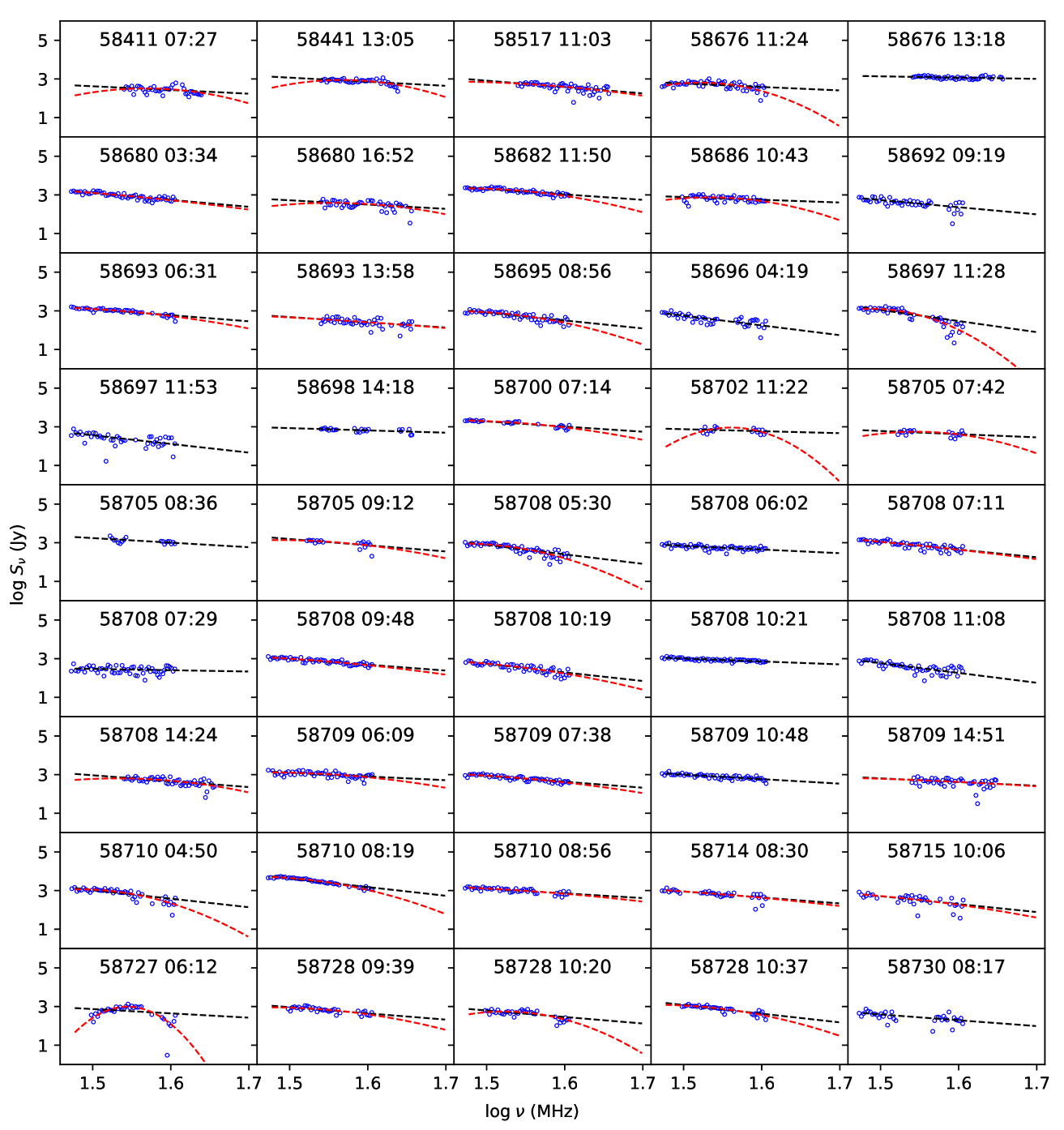}
 \caption{The log-log plots of MRA spectra fitted with power law and log-normal functions. The black line denotes the power law fit and the red line denotes the log-normal fit. The log-normal fitting parameters are not available for some spectra and only the power law fitting is shown for those cases. Each subplot data is labelled with the MJD day and UTC hour at the time of occurrence. The frequency axis goes from 30--50 MHz.}
 \label{spec1}
\end{figure}

\begin{figure}[h!]
\centering
 \includegraphics[width=\textwidth,height=\textheight,keepaspectratio]{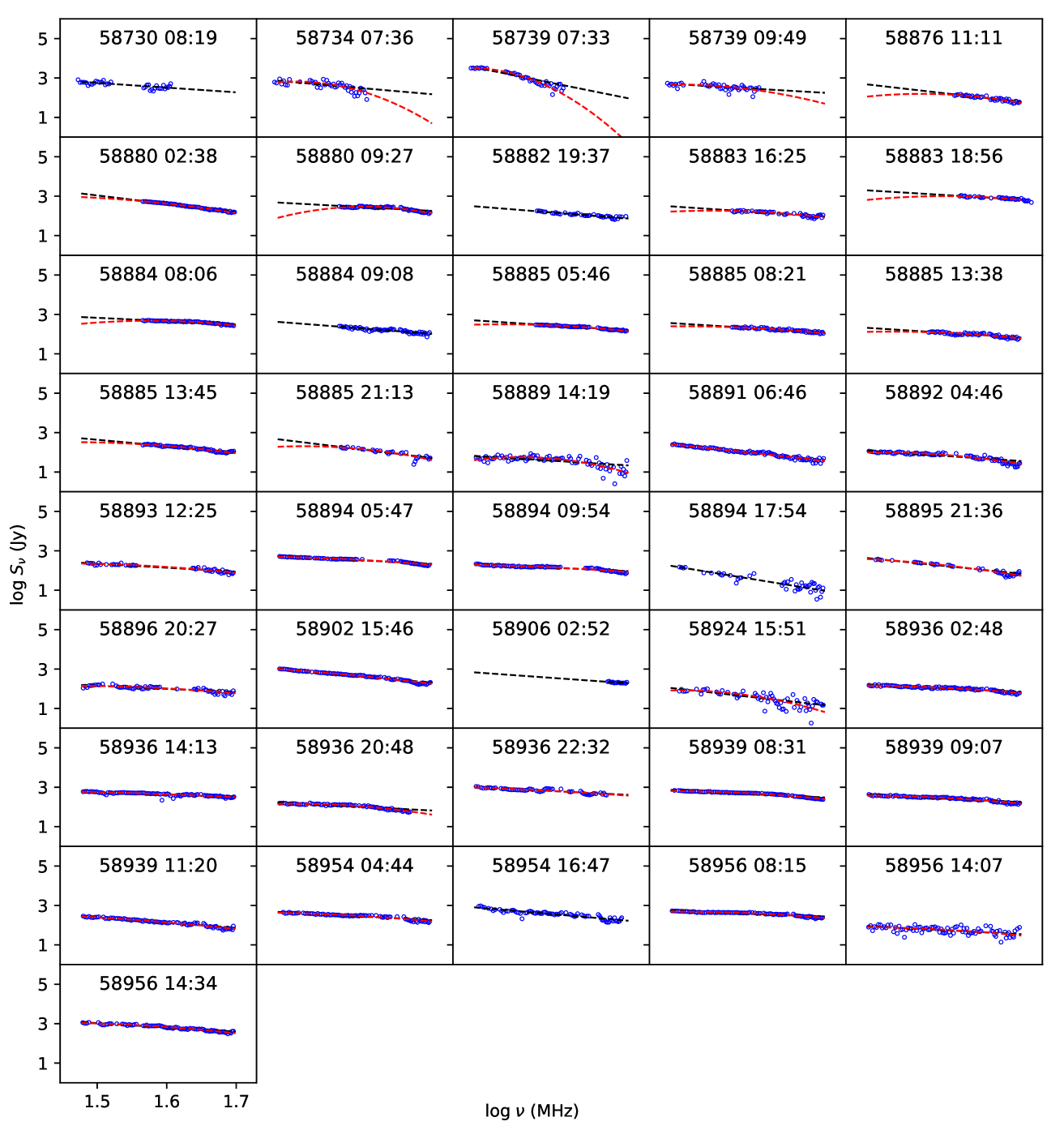}
 \caption{The log-log plots of MRA spectra fitted with power law and log-normal functions. The black line denotes the power law fit and the red line denotes the log-normal fit. The log-normal fitting parameters are not available for some spectra and only power law fitting is shown for those cases. Each subplot data is labelled with the MJD day and UTC hour at the time of occurrence. The frequency axis spans from 30--50 MHz.}
 \label{spec2}
\end{figure}

We collected the spectra of 86 MRAs (49 MRAs with 10 MHz bandwidth, and 10 s integration and 37 MRAS with 20 MHz bandwidth and 5 s integration) and calibrated them using Cyg A. The calibrated spectra were averaged over 200 kHz bandwidth to improve the per-channel SNR ratio. The averaged spectra of 86 MRAs fit with both power law and log-normal functions are shown in Figure \ref{spec1} and \ref{spec2}. The MJD date and UTC time of the events are labelled in each subplot.
The Orville imager in the first four months of operation experienced some technical issues limiting it from achieving the expected sensitivity with 10 MHz bandwidth. This has increased the noise per channel for the MRAs before MJD 58740 which is also evident in the spectra (see Figures \ref{spec1} and \ref{spec2}).
A spectral index was derived for each spectrum from the power law fit (see Section 1). The measured spectral indices of MRAs varied from --0.650 to --7.106.

The missing data in some spectra are primarily due to occasional ADP server dropouts (e.g. event from MJD 58700, 58702 in Figure 3) and flagging of bad RFI channels (e.g. the event from MJD 58892 in Figure 4).
Each ADP server is configured to output 1.8 MHz data while running at 10 MHz bandwidth and  3.3 MHz of data at 20 MHz bandwidth. The ADP failures resulted in the loss of data chunks from certain channels which is evident in some spectra. Most of  the MRAs follow a power law dependence on frequency having higher flux density at lower frequency. Most of the spectra are smooth but some have wiggles through out. Some spectra even have unexpected bumps which could be possibly due to leakage of bright narrow RFI into nearby channels or could be intrinsic features of the spectra.

As can be seen from the spectra, particularly those in Figure \ref{spec2} with 20 MHz bandwidth,  many of the spectra  deviate from a power law fit, which would appear as a straight line. Many events contain curvature, and some even appear to have a spectral turnover. For these reasons, we searched for a more appropriate model to fit the spectra. One promising candidate is a log-normal distribution, where the spectral power can be fit as a Gaussian with respect to the logarithm of frequency. The unnormalized form of the log-normal distribution can be written as 
\begin{equation}
    S_{\nu} = Ae^{-\frac{(\log(\nu) - \log(\nu_{0}))^{2}}{2\sigma^{2}}}
\end{equation}
where $S_{\nu}$ is the flux density,  $A$ is the scale factor, $\nu$ is the frequency, $\nu_{0}$ is the turnover frequency and $\sigma$ is the standard deviation of the log-normal distribution.

\begin{figure}[htb]
\centering

    \includegraphics[height = 0.4\textwidth, width=0.95\textwidth]{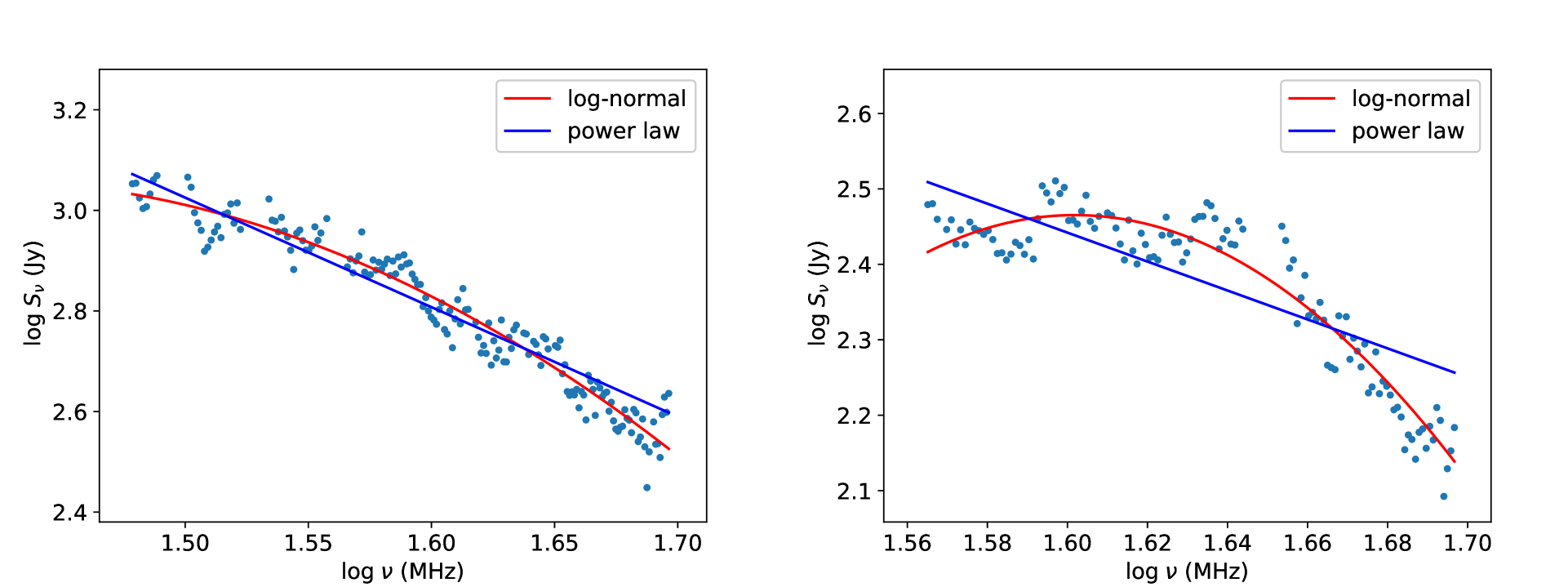}

  \caption{Comparison between the power law and log-normal fit of 2 MRA spectra. (Left) Event from MJD 58956 UTC 14:34 without spectral turnover in the log-normal space. The AIC score for the log-normal fit is 1331 and for the power law fit is 1392. (Right) Event from MJD 58880 UTC 09:27 with a spectral turnover in the log-normal space. The AIC score for the log-normal fit is 703 and for the power law fit is 831. The small scale sinusoidal variations are intrinsic features of each spectra. Neither fitting method can account for the variations of these intrinsic features at such a small spectral scales.}
  \label{comp}
\end{figure}

 In order to understand which fitting scheme works best, power law or log-normal function, we used the Akaike Information Criterion (AIC) \citep{ak74}. The AIC score predicts the quality of a model relative to other models. The AIC criteria takes into account of the goodness of fitting and the complexity of the model. In this method, the AIC score for different models are calculated and the model with the lowest AIC score is chosen.  The AIC is a relative metric in understanding which model fits the data better and the value itself cannot measure the goodness of the fit. The AIC score is given by 
 \begin{equation}
     AIC = 2k - 2\log (L)
 \end{equation}
  where $k$ is the number of model parameters and $L$ is the maximum value of the likelihood function for the model. In the case of least square regression with normally distributed errors, AIC can be calculated by
  \begin{equation}
      AIC = 2k + n\log\left(\frac{\sigma}{n}\right)
  \end{equation}
  where $k$ is the number of model parameters, $n$ is the number of observations and $\sigma$ is the residual sum of squares. The model parameters of the power law and log-normal fitting are derived using least square regression analysis and equation 4 can be used to calculate the AIC score.
   Figure \ref{comp} shows a comparison between the log-normal and a power law fit for two separate events, one of which contains a turnover. For the first event without turnover, the log-normal AIC score is 1331 and the power law AIC score is 1392. In the second event with turnover, the log-normal AIC score is 703 and the power law AIC score is 831. In both cases, the log-normal fitting has the lower AIC score and fit the spectra better than the power law which is evident from Figure \ref{comp}. We note that the improvement gained by the log-normal fit may just be a product of the fact that a log-normal distribution is a function of three parameters, whereas the power-law has only two. With one more degree of freedom the log-normal fit may just be more adept at handling a more complex structure. However, it also may be that the log-normal fit is a better approximation of the physics behind the emission mechanism. 
The turnover frequency varies from source to source depending on the emission mechanism and absorption along the line-of-sight. Certainly a power-law could not hold up for all frequencies, as this would result in a runaway at low frequencies.  Therefore, it makes sense to use a function that allows for a spectral turnover, as the log-normal does.
However, the log-normal fitting fails if the fitting method cannot settle on a reasonable turnover frequency value below 30 MHz. This is mainly due to the presence of strong RFI in the data and low SNR for some events before MJD 58740 due to the technical issues with the correlator. Therefore, while power law fitting of the spectra was carried out for all 86 MRAs the log-normal fitting was only carried out for 67 MRAs. Even though, both the fitting methods works well for mapping out the overall spectral shape, neither of the methods account for the small scale spectral variations. The spectra of two events from Figure \ref{comp} shows small scale sinusoidal variations which are not captured by the two fitting methods. The scintillation caused by the E region could be one of the reasons causing this variation. In that case, we should be seeing such variations more during the day time events. However, such sinusoidal spectral variations have been observed during the day time as well as night time. Therefore, it is highly likely that these could be small scale intrinsic spectral structures of MRAs. The next section will describe the statistical analysis of the spectral parameters from each fitting method and their comparison with the physical properties of MRAs.

\section{Statistical Analysis of MRAs}
\begin{figure}[h!]
\centering
\includegraphics[height = 11cm, width=0.95\textwidth]{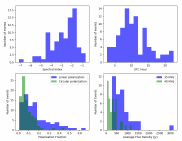}
 \caption{Histogram showing the distribution of spectral index, time of MRAs in UTC, polarization fraction and average flux density of MRAs.}
 \label{multi_hist}
\end{figure}

We conducted statistical analysis to search for correlations between the spectral parameters and the physical properties of MRAs. The derived spectral parameters include the spectral index from the power law fitting and the turnover frequency and standard deviation from the log-normal fitting. Figure \ref{multi_hist} shows the distribution of the spectral index, UTC time of occurrence, linear and circular polarization fraction and the average flux density of MRAs. The top left panel of Figure \ref{multi_hist} shows the histogram of the spectral index distribution. The spectral index distribution peaks around --1.73 suggesting that the majority of the spectra are not extremely steep as those reported in \citet{ob15b,ob16a}. The top right panel of Figure \ref{multi_hist} shows the UTC time of occurrence of MRAs peaking between 8 and 12 UTC. This fits with expectations since meteors usually have peak occurrence between 02.00 to 06.00 AM local time. 
The bottom left panel of Figure \ref{multi_hist} shows the histogram distribution of the linear and circular polarization fraction measured for MRAs. The circular polarization fraction peaks around 0.05 and linear polarization fraction peaks around 0.07. Some of the events have higher linear polarization while most of the events have lower circular polarization fraction. Previous work by \citet{ob15b} has measured high amounts of broadband linear polarization in one of the beamformed observations though such events are rare.  The bottom right panel of Figure \ref{multi_hist} shows the average flux density measured for MRAs at a center frequency of 35 and 40 MHz, showing a higher fraction of events towards the low flux density region.  This measurement is also in agreement with the flux density distribution from \citet{ob16a}. The Orville imager has been able to detect faint MRAs with a lowest measured flux density of 41 Jy at 40 MHz.  

Using the radio data,  we measured features like flux density, angular size, etc.\ which have observational biases. Calculating the corresponding physical parameters like luminosity, physical size, etc., requires distance information. Most of the MRAs presented in this paper were observed only with LWA-SV and their distance information is not available. In order to calculate the altitude and distance of MRAs observed from LWA-SV, we used the optical data from the Global Meteor Network (\url{https://globalmeteornetwork.org}). The optical data contained the beginning and ending of the optical trajectory in latitude, longitude and altitude units. These values were used to create a series of trajectory points in the  ECEF (Earth centered Earth fixed) coordinate system. We looked for optical events happened within 30 seconds of the MRA. From LWA-SV, we know the azimuth and elevation of the MRA. Triangulation using the angular direction of the MRA from LWA-SV and the optical trajectory coordinates will give us an estimated value of the radio emission along the trajectory of the optical trail. The coordinates of radio emission in ECEF is then converted back to latitude, longitude and altitude units. The triangulation fails if the angular direction of MRA is outside the optical trajectory. Finally, the distance between the MRA and LWA-SV is estimated for those events. Out of 86 events, optical data was only available for 28 events. The rest of the events occurred either during the day or at times when the optical data were not available. Altitudes and distances of 4 other MRAs were calculated using triangulation from the two LWA stations. Most of the MRAs that occurred are not seen by LWA1 due to the lower sensitivity of LASI or because LWA1 was not generating all-sky images. This resulted in obtaining altitudes and distances for 32 MRAs only. \citet{ob16b} measured a typical height of 100 km for MRAs and we used that value as the altitude for events which did not have optical data or radio data from LWA1. For the analysis presented in this paper, we calculated the average luminosity (product of average flux density and distance squared), physical size (product of angular size from image and distance), average energy (product of average luminosity and FWHM of the duration peak), incidence angle (angle between the meteor trajectory and tangent to Earth's surface only for events with altitude information). Along with that, we also utilize the velocity and mass of the optical meteors from the Global Meteor Network to calculate the kinetic energy of the parent meteoroid producing MRAs.

Depending on the radiant direction of the optical meteors, the Global Meteor Network have classified meteors as sporadic and shower ones. Also, each of the shower meteors have been associated with their corresponding shower label. Out of 28 MRAs with optical data, 15 MRAs originate from known meteor showers (8 from Perseid meteor shower) and 13 MRAs originate from sporadic meteors. Therefore, at least from this small sample of optical data, MRAs can form from  shower meteors as well as sporadic meteors. Further combined radio and optical observations in future are required to better understand the sporadic or shower origin of MRAs.
\begin{figure}[h!]
\centering

    \includegraphics[width=.95\textwidth]{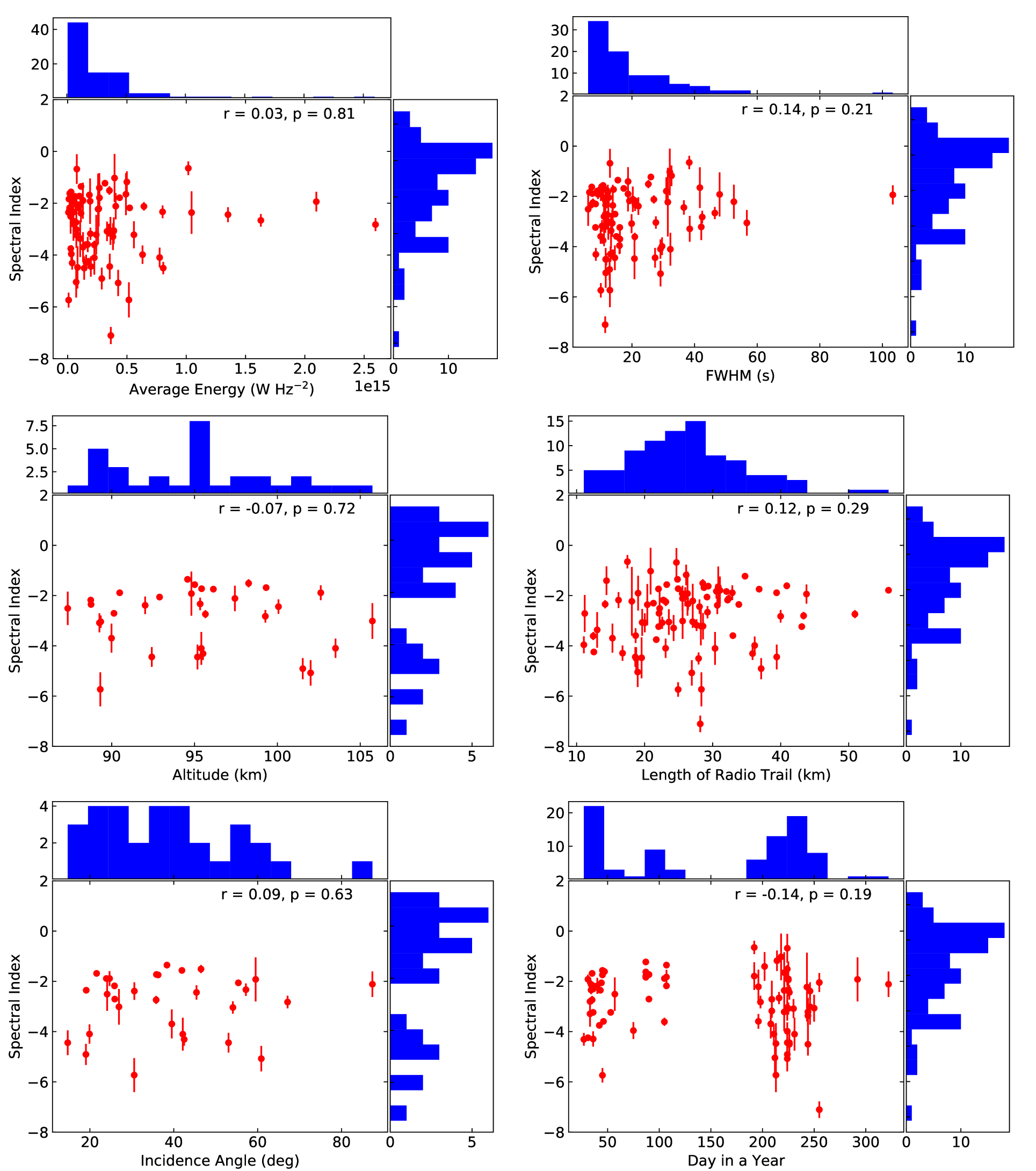} 
  \caption{Scatter plots with histograms on the sides showing the comparison between spectral index and different physical properties of MRAs. The calculated Pearson correlation coefficient (r) and two-tailed p value are labelled in each subplot. }
  \label{multi_par1}
\end{figure}

Figures \ref{multi_par1} shows the scatter plots comparing the spectral index with different physical properties of MRAs with their histogram on the sides. 
Each subplot is labelled with Pearson correlation coefficient (r) and two-tailed p value for non-correlation calculated using the Python Scipy package ({\tt scipy.stats.pearsonr}). The r value varies from +1 to --1, where a +1 indicates a maximum positive correlation, --1 indicates a maximum negative correlation and 0 indicating no correlation. A p value less than 0.05 suggest that the correlation is significant.  Hence, a higher absolute value of r and a lower p value indicates strong correlation. Figure \ref{multi_par1} shows the plots of spectral index with average energy, FWHM, altitude, length of radio trail, incidence angle and day in a year.  In the  plot between spectral index and average energy, events with higher energy tend to be flatter ($|\alpha| < $ 1).   A similar trend is also observed in the plot between spectral index and FWHM. The longer duration events tend to be flatter. However, the r and p value suggest that the correlation may not be significant. More events with longer durations are needed to examine the flatter spectral trend at longer FWHM. The plots of spectral index with the altitude, length of radio trail and incidence angle do not show any obvious correlations.
The histogram of the length of radio trail distribution peaks around 25 km. Also, the plot between the spectral index and day in a year do not show any seasonal correlation. However, the data from the middle of the year was not collected and only a few events were collected at the end of the year. 

\begin{figure}[h!]
\centering
    \includegraphics[height = 0.41\textwidth,width=.95\textwidth]{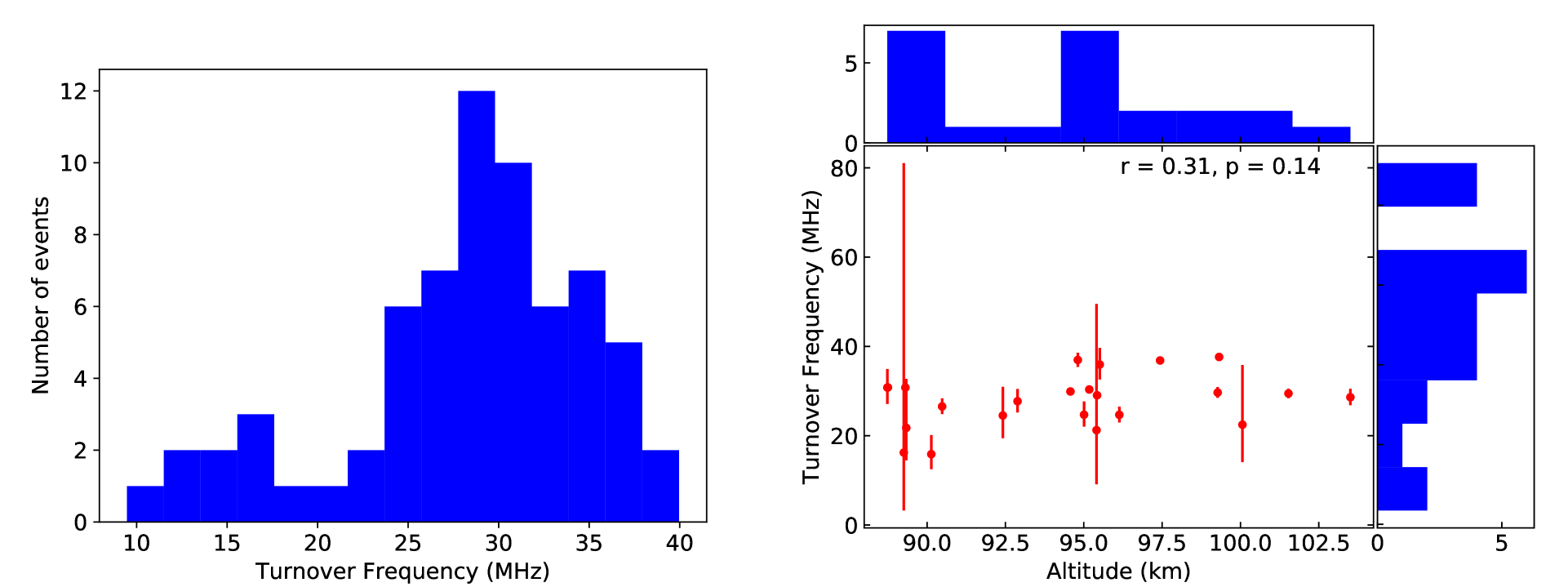}

  \caption{Histogram plot of the turnover frequency and scatter plot with histograms on the sides showing the comparison between the turnover frequency and altitude of MRAs. The calculated Pearson correlation coefficient (r) and two-tailed p value are labelled in the second subplot. }
  \label{multi_hist_logn}
\end{figure}

Top left panel of Figure \ref{multi_hist_logn} shows the distribution of the turnover frequency from the log-normal fitting of the MRA spectra. The MRA observations start from 30 MHz and the turnover frequency distribution peaks around 29 MHz suggesting that most of the MRAs did not undergo log-normal turnover. At the same time, 42\% of MRAs have undergone log-normal turnover between 30--40 MHz. 

If MRAs are caused by plasma radiating at the local plasma frequency, then the  spectrum of an MRA gives a measure of the amount of plasma that is emitting in addition to its radiation efficiency at each frequency. As the plasma diffuses into the background, the plasma from the dense meteor trail core comes into equilibrium with the low density background. As a result, the majority of the plasma is at the background density, and a small amount is at the initial peak density. The amount of plasma radiating near the plasma frequencies are proportional to the square root of the local electron density. Therefore, during this process, there will be a lower amount of high density plasma emitting at high plasma frequencies and higher amount of low density plasma emitting at low frequencies. This explains the trend of decreasing luminosity as frequency increases above $\sim$ 30 MHz. The downward trending curve (frown-shaped) observed for many events indicate a decrease in luminosity at very low frequencies as well as at very high frequencies. This may be caused by the absorption of low frequency radio waves due to electron-neutral collisions.   Finally, the efficiency of the radiation likely varies with size scale (i.e. wavelength), and it is likely that the observed spectra could be the result of these three factors.

The top right panel of Figure \ref{multi_hist_logn} shows the comparison between turnover frequency and altitude. The turnover frequency shows a weak positive correlation with the altitude.  With decreasing absorption at higher altitudes we would expect lower frequencies to be brighter, leading to lower turnover frequencies at high altitude. However, this is exactly the opposite of what is observed and it remains a mystery. The higher p-value possibly suggest that this correlation may not be significant as well.

\citet{dy11} have conducted simulation studies to understand the global variation of meteor trail plasma turbulence. In this work, they focused on the duration of the plasma instabilities and the various factors affecting it. The study utilized a meteor trail evolution model from entry to diffusion in the atmosphere. The model was used to simulate head echoes and nonspecular echoes for an incoming meteoroid with a specific mass, velocity and composition values. The duration of plasma instabilities were measured using the duration of nonspecular echoes as they are closely associated. For more information on plasma formation around the meteoroid and plasma turbulence of non-specular echoes, see \citet{sug18,yee13}. \citet{dy11} showed that the turbulence duration is primarily affected by the background electron density, neutral winds or drifts, mass and velocity of the meteoroid. They predicted longer turbulence duration during night time as the electron density is 2 orders of magnitude less compared to the day time. Similarly, they predict longer duration with increasing meteoroid mass and during high neutral wind speeds. The velocity of the meteoroid also affects the turbulence duration, however the relation is highly complicated. Theoretical and simulation studies conducted by \citet{di06a,di06b} shows that background plasma density is an important factor affecting the meteor diffusion unless the meteor trail density is greater than the background density by 3 orders of magnitude. The study predicts that the instability growth rate is higher during night time compared to day time causing difference in the diffusion rates. Also, the observation studies conducted by \citet{op08} have shown that meteors tend to produce more and longer duration non-specular echoes during night time compared to day time.

\begin{figure}[h!]
\centering

    \includegraphics[width=.97\textwidth]{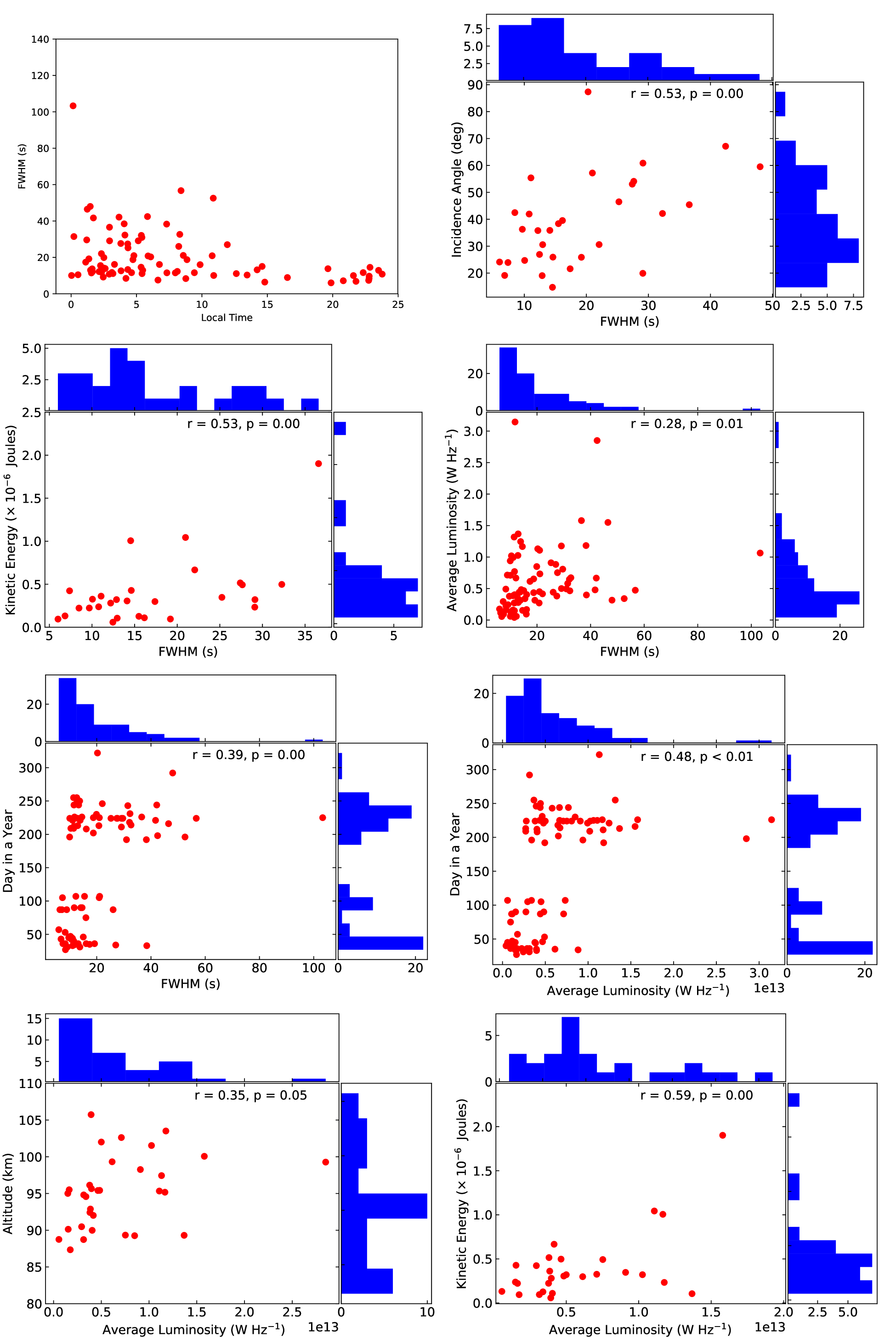}   
    
      
  \caption{Scatter plots with histograms on the sides showing the comparison between the different measured physical properties of MRAs. The calculated Pearson correlation coefficient (r) and two-tailed p value are labelled in each subplot. }
  \label{multi_par2}
\end{figure}

 In the case of MRAs, both the current emission mechanism hypotheses, plasma wave and transition radiation mechanisms requires plasma instabilities for longer timescales compared to the ablation timescales. The FWHM duration of MRAs is possibly a measure of these plasma instability timescales. The first 5 subplots of Figure \ref{multi_par2} show the comparison between different physical properties and the FWHM duration of MRAs. The plot between FWHM and local time suggests that the duration of MRAs is longer during night time and it decreases during day time. This is in agreement with the predictions from the \citet{dy11}.

In the next subplots, we observe positive correlation of the FWHM with incidence angle, kinetic energy of the meteoroid, luminosity and day in a year. The correlation of FWHM with incidence angle may be related to how the incoming meteor ablates in the atmosphere. At this point it is not clear why we see a correlation between the incidence angle and FWHM. Also, previous radar/radio observations have not reported any correlations between the same quantities. The ionization efficiency of the meteoroid increases with the velocity \citep{haw56}  and changes with composition. Therefore, the higher the kinetic energy of the meteoroid, the higher will be the ionization efficiency producing turbulent plasma for longer duration. The radio and optical observations presented in the paper do not provide composition information. Future studies will investigate the variation of the physical properties as a function of the meteoroid composition. 


Also, the events with higher luminosities arise from longer duration plasma instabilities. Third panel of the  Figure \ref{multi_par2} shows that the day in a year is correlated with the FWHM and average luminosity. The events between July and September have higher FWHM compared to the events between January and April. Since luminosity is correlated with FWHM, a similar trend is also observed for the plot between day in a year and average luminosity. As we discussed in Section 5, the instrument was having sensitivity issues for the events between July and December. This may be the reason we do not find many events with lower luminosities after July. But it still does not explain why many events have higher FWHM and luminosities at the same time. This possibly suggests that they are in fact a different population of meteoroids compared to the ones in the beginning of the year. Many events between July and September are from the Perseid meteor shower and a combination of above discussed factors may result in higher FWHM and luminosities.

The bottom panel of Figure \ref{multi_par2} shows the positive correlation of luminosity with the altitude and kinetic energy. The correlation between luminosity and altitude could be explained by the decrease in absorption at higher altitudes since the neutral density decreases as the altitude increases. Similarly, the mean free path of an electron increases with altitude, allowing it to radiate more energy before colliding with a neutral. Both the kinetic energy of the meteoroid and the luminosity depends on FWHM and they are correlated to each other. Figure \ref{scheme} summarizes the correlation of FWHM with the local time (perhaps ionospheric electron density), incidence angle, luminosity and kinetic energy and similarly the correlation of luminosity with kinetic energy and altitude.

\begin{figure}[h!]
\centering
    \includegraphics[width=0.5\textwidth]{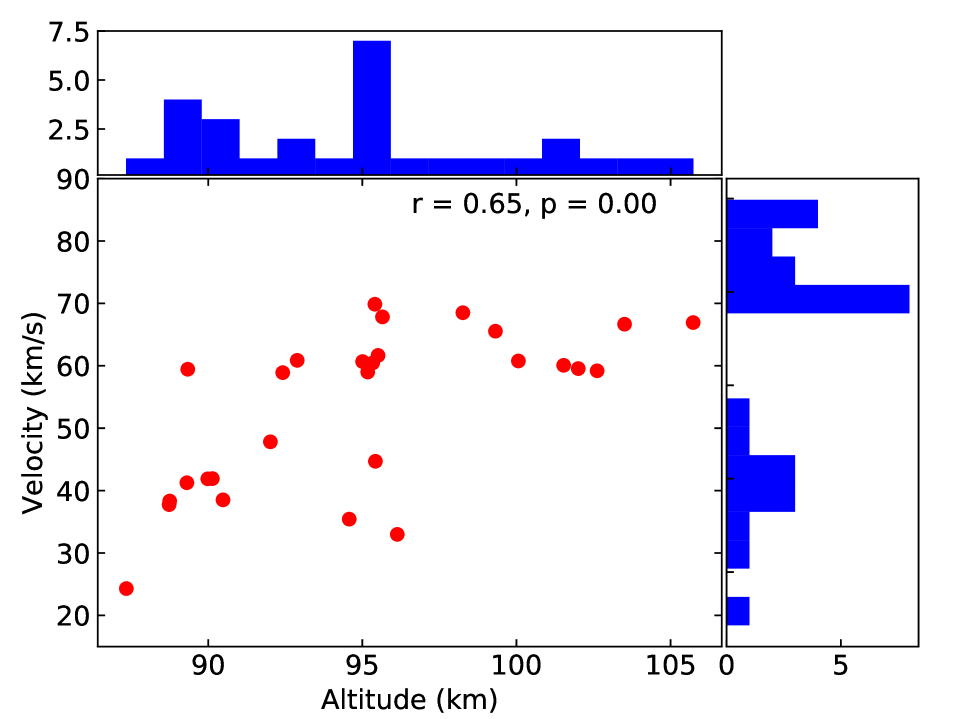}   
  \caption{Scatter plots with histograms on the sides showing the comparison between the velocity of optical meteor and altitude of MRAs. The calculated Pearson correlation coefficient (r) and two-tailed p value are also labelled. }
  \label{vel_alt}
\end{figure}

The ionization efficiency of meteoroids changes as a function of velocity. Similarly, \citet{cep98} have shown that the height of maximum ionization scales with the velocity of meteors. MRA formation requires long duration plasma instability and has a higher probability to form at heights having maximum ionization of the incoming meteoroid. Figure \ref{vel_alt} compares the velocity of the optical meteors with the altitude of MRAs. From the plot it is clear that the higher velocity meteors tend to form MRAs at higher altitudes which agrees with our prediction.

It is not immediately clear why spectral parameters obtained from the two fitting methods do not correlate with the physical properties of MRAs. At the same time, the FWHM and the luminosity of the MRAs are correlated with other physical properties. Even though spectral index tends to be flatter for longer duration events in the current analysis, more observations of events with longer FWHM are required to know for certain. The non-correlation of spectral parameters with the physical properties suggest that the energy distribution does not depend directly on the different physical properties presented in the paper. However, the correlation of FWHM and luminosity with some physical properties indicate that these properties play a significant role in driving the turbulent plasma instability required for the formation of MRAs. In the case of transition radiation \cite{plat02}, the radiation mechanism relies on the plasma structure. Therefore, it is possible that the spectra of MRAs might depend on the turbulent structure of the plasma. Further studies are needed in future to understand various factors affecting the turbulent structure of the plasma and how that affect the spectra of MRAs. However at this point, calculating this complicated dependence is beyond the scope of this paper.

\begin{figure}[h!]
\centering
\includegraphics[width=11cm]{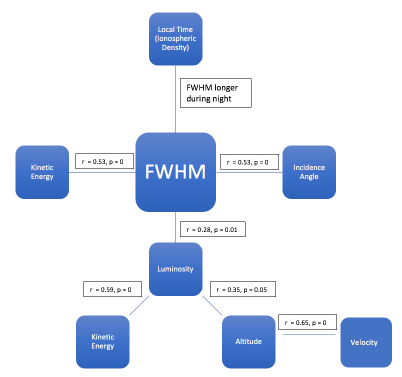}
 \caption{Schematic diagram showing the correlation between different physical properties and their corresponding correlation values.}
 \label{scheme}
\end{figure}

\subsection{Principal Component Analysis (PCA)}
In this section, we use the principal component analysis (PCA) to search for unknown correlations within the MRA data set. PCA can be generally used to perform feature dimensionality reduction where the input data is projected on to a lower dimensional surface maximizing the variance of the data. This helps to interpret the data without losing much information (see review by \citet{jol16} and the references therein). In this method, a set of possible correlated features are converted to a set of new orthogonal, uncorrelated variables. These new variables are linear functions of the original features and are generated by maximizing the variance of the data. These new variables are known as the principal components. Maximum variance of the data will be in the first component and lesser in the successive components. Therefore, looking at the first few components will help the user to visualise the high dimensional data set as a lower dimensional projected data set. This is widely used in machine learning to reduce data storage, speeding up learning algorithms and for the visualization of the data. Genetic algorithms and machine learning have been used previously to extract meteoroid parameters \citep{roy09,tar19}. For this purpose, we used the PCA module ({\tt sklearn.decomposition.PCA}) from the {\tt scikit-learn} machine learning tool in Python \citep{scikit-learn}. Using PCA, we can study the features causing maximum variance in the first few components or how much of the data is explained by each component.



The PCA was conducted with spectral parameters derived from the two fitting methods and the measured physical properties of MRAs. The spectral parameters from the two fitting methods were available for 67 MRAs. We used 14 measured features for PCA and they are (1) UTC hour of occurrence in a day, (2) average luminosity, (3) average energy, (4) physical size, (5) day in a year, (6) altitude, (7) incidence angle (8) linear polarization fraction, (9) circular polarization fraction, (10) spectral index, (11) turnover frequency, (12) log-normal standard deviation, (13) FWHM in time (14) kinetic energy (calculated using mass and velocity) of the parent meteoroid. Since features like altitude, velocity, mass and incidence angle were not available for all the events, PCA was done in two sets of data as it cannot handle missing values. In the first set (DS1), we used all the features of 67 MRAs except the altitude and incidence angle info. In the second set (DS2), PCA was performed using 19 MRAs which had  altitude, velocity and incidence angle information.

Before feeding the feature set to the PCA algorithm, all features were standardized to zero mean and unit variance. The standardized data set was fed into the PCA algorithm to obtain the principal components. The algorithm  outputs principal components as a linear combination of the original feature set and the percentage of the variance calculated for each principal component. 

\begin{figure}[h]
\centering

  \includegraphics[ width=0.95\textwidth]{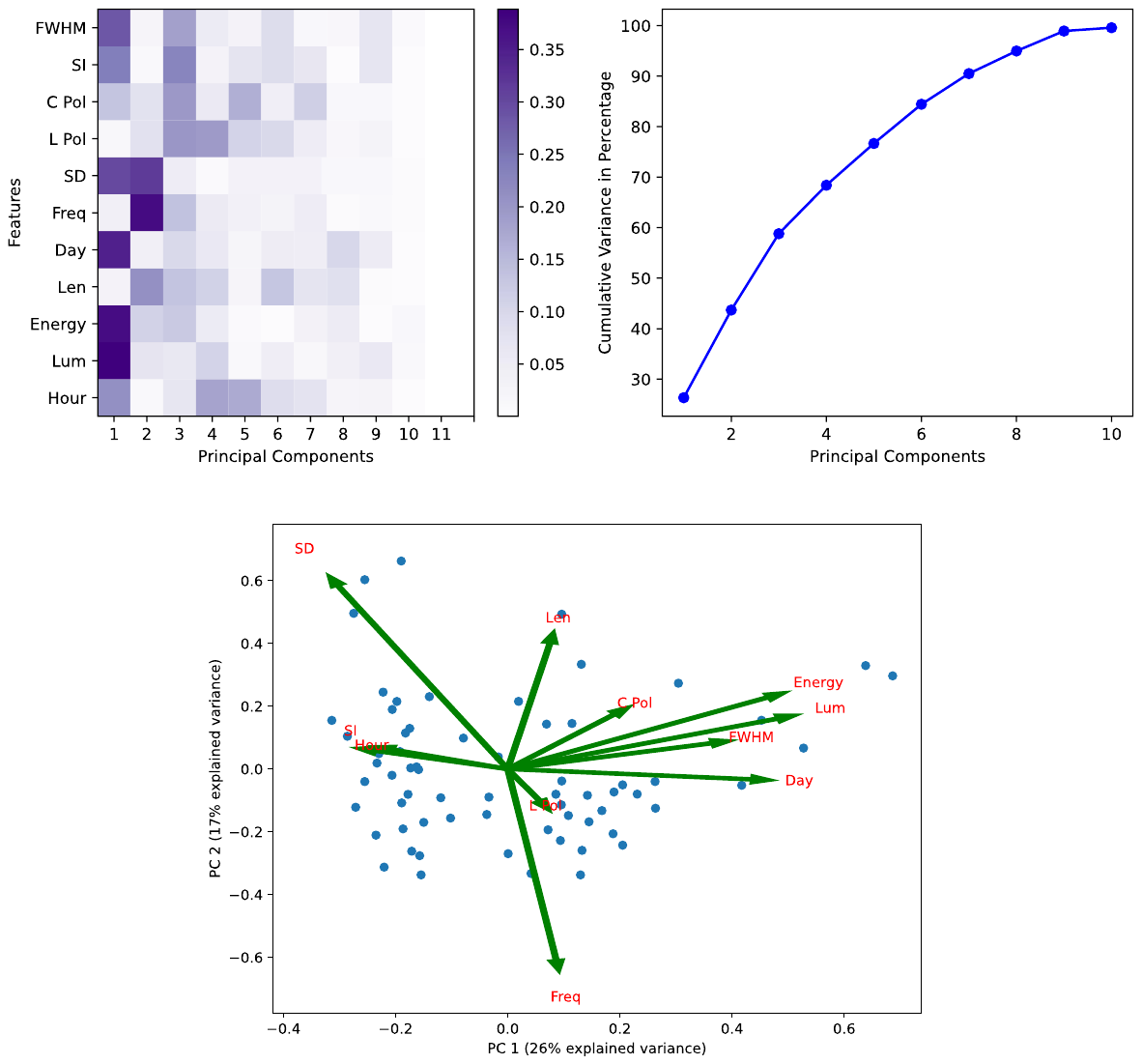}
  \caption{(Top left) Plot shows the color map of principal components from the PCA analysis with the features from DS1. The features on y axis are H : hour of occurrence, Lum : average luminosity, Energy : average energy, Len : length of radio trail, day : day in a year, Freq : turnover frequency, SD : standard deviation, L Pol : linear polarization fraction, C Pol : circular polarization fraction, SI : spectral index and FWHM : full width at half maximum duration of the event. (Top right) plot shows the cumulative variance as a function of the number of principal components. (Bottom) Biplot from the PCA on DS1 showing the first two principal components and how each feature affect the components. The arrow denotes the feature vector in the principal component space. } 
  \label{multi_pca}
\end{figure}

Figure \ref{multi_pca} shows the results of PCA from DS1.
  The top left panel of  Figure \ref{multi_pca} shows the principal component vector where the color map denotes the contribution of each feature to the principal components. For visualization, component vectors are standardized (zero mean and unit variance). The absolute values of vectors are multiplied by the variance explained by each component. In the first component, luminosity, energy, day, standard deviation and FWHM capture the maximum variation. The second and third principal components suggests that the turnover frequency and spectral index also contribute to the variance.
  The top right panel of Figure \ref{multi_pca} shows the cumulative variance as a function of principal components. The first 4 components explain the 69\% of the data. Variance is maximum for the first component ($\sim$26\%) and decreases for successive components.

Bottom panel of Figure \ref{multi_pca} shows the PCA biplot from DS1 with the first two principal components and feature vectors in the principal component space. The scatter plot has the first two principal components explaining roughly 43\% variance of the data. The arrows in the plot represent the feature vector in the principal component space. The projection of the vector explains the contribution of features towards each components. This plot can be used to understand the correlation between different features. The cosine of the angle between the feature vectors is proportional to the correlation coefficient between the features \citep{jol16}. The smaller the angle between two features, the greater the positive correlation between features. An angle of 180 degrees implies a negative correlation between the features. The length of the vector is proportional to the contribution of each feature to the principal components. 

From the bottom panel of Figure \ref{multi_pca}, we see that there is a positive correlation between energy and FWHM and between luminosity and day in a year. Also, there is a weak correlation between the spectral index and hour of occurrence. The spectral index and turnover frequency do not show any significant correlation with other features.

\begin{figure}[h]
\centering

\includegraphics[width=0.95\textwidth]{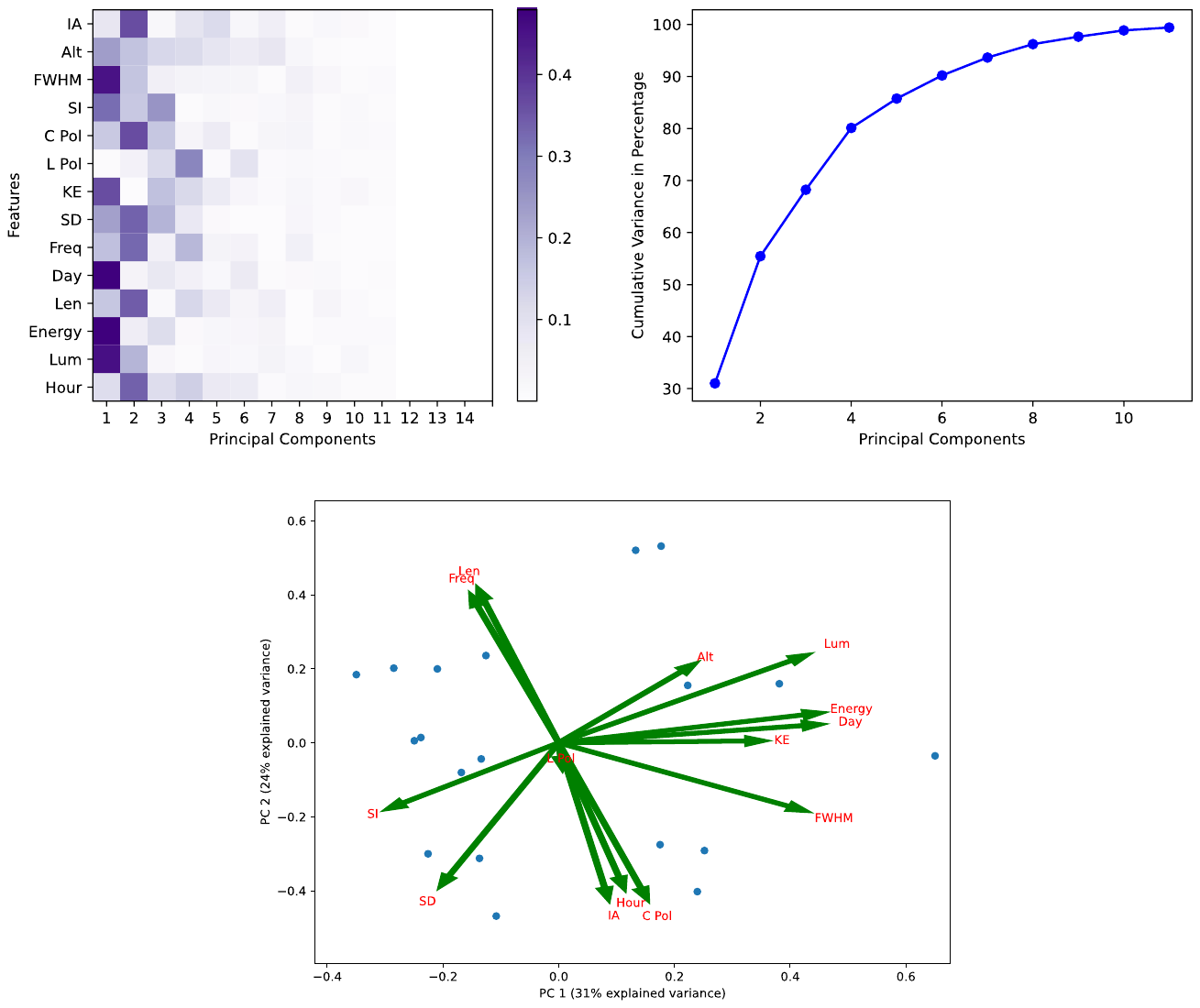}
  \caption{(Top left) Plot shows the color map of principal components from the PCA with the features from DS2. The features on y axis are H : hour of occurrence, Lum : average luminosity, Energy : average energy, Len : length of radio trail, day : day in a year, Freq : turnover frequency, SD : standard deviation, KE : kinetic energy of the parent meteoroid, L Pol : linear polarization fraction, C Pol : circular polarization fraction, SI : spectral index, FWHM : full width at half maximum duration of the event, Alt : altitude and IA : incidence angle . (Top right) Plot shows the cumulative variance as a function of the number of principal components. (Bottom) Biplot from the PCA on DS2 showing the first two principal components and how each feature affect the components. The arrow denotes the feature vector in the principal component space.  }
  \label{multi_pca_2}
\end{figure}

Figure \ref{multi_pca_2} shows the results of PCA from DS2.
  The top left panel of  Figure \ref{multi_pca_2} shows the  contribution of each feature to the principal components. In the first component, luminosity, energy, day, kinetic energy, spectral index and FWHM capture the maximum variation. The second principal component suggests that the hour of occurrence, length of radio trail, turnover frequency, standard deviation, circular polarization and incidence angle also contribute to the variance. From the top right panel of Figure \ref{multi_pca_2}, the first 4 components explain the 79\% of the data. Variance is maximum for the first component ($\sim$31\%) and decreases for successive components.

The bottom panel of Figure \ref{multi_pca_2} shows the PCA biplot from DS2 with the first two principal components and feature vectors in the principal component space. The scatter plot has the first two principal components explaining roughly 55\% variance of the data. The biplot shows a positive correlation between altitude and luminosity. The turnover frequency shows weak correlation with the length of the radio trail. Both the turnover frequency and length of radio trail information are available for 67 events. But the biplot from Figure \ref{multi_pca} shows that these two features are uncorrelated. More importantly, the FWHM seems to be correlated with luminosity, kinetic energy, day in a year, energy, incidence angle, circular polarization and hour of occurrence. At the same time, the spectral index does not show any strong correlation with other physical properties of MRAs.


\section{Conclusions}

In this work, we presented the spectra of 86 MRAs using the new broadband imager for LWA-SV and conducted a statistical analysis of the measured spectral and physical properties for each event. The spectra of MRAs were fit with both power law and log-normal functions. The spectral parameters derived from each fitting method was compared with physical properties of MRAs. The measured spectra are mostly smooth and usually follow a power law with frequency getting brighter at lower frequencies within the observed frequency range. However, some MRAs have spectral turnovers at low frequencies (see Figure \ref{comp}).  The spectral index distribution of the MRAs peaked at  --1.73. This value is much flatter than the previous spectral index measurements by Obenberger et al.. 

This could be possibly due to two reasons, either the small number of MRAs  collected  in Obenberger et al.  were unusually steep or the lack of astrophysical calibration of the same observations could have introduced an error in the spectral curvature. Also, the --1.73 spectral index value is much flatter than the --3.7 spectral index upper limit reported by \citet{zha18}. With an average spectral index value of --1.73, MWA should have detected many MRAs within their frequency range. But the survey conducted by \citet{zha18} did not detect any MRAs at higher frequencies between 72--103 MHz. This suggests that the power law approximation for the spectra of MRAs do not extend out to frequencies higher than $\sim$ 60 MHz.

Even though radio source spectra are typically characterized by power law, we note that a log-normal distribution provides a better fit to the spectra presented in this paper. The spectra of many MRAs undergo a log-normal turnover between 30-40 MHz. The UTC hour of occurrence of MRAs peaks between 8--12 UTC or 2--6 am local time at which meteors peak as expected for MRAs. 
The distribution of the polarization fraction and flux density are in agreement with the previous observations \citep{ob14b,ob16a}. We find that the distribution of length of MRA trails peaks at 27 km. 
 
Results from the statistical analysis and PCA shows that the spectral parameters derived using two fitting methods do not show any strong correlation with the physical properties presented in this paper. The log-normal turnover frequency shows a weak correlation with the altitude, but it is based on a small number of sources. Similarly, the spectral index derived from the power law fits do not show any strong correlation with measured physical properties of MRAs. Even though no apparent correlations have been observed between spectral parameters and physical properties, we observe several interesting correlations among the physical properties. The FWHM duration of MRAs are correlated with the local time (perhaps ionospheric density), incidence angle, luminosity of MRAs and kinetic energy of the parent meteoroid. Similarly, the luminosity of MRAs is correlated with the altitude and kinetic energy of the parent meteoroid as well. Also, the velocity of optical meteors are highly correlated with the altitude of MRAs. This work updates the measured spectral parameters of MRAs from a large sample population using two fitting methods. In the future, we need to lower the frequency range of observations to get down to the low 20s of MHz to better measure the spectral turnovers. Also, the observed correlations within this study can be used to constrain theoretical models of MRAs and to study their emission mechanism.

%
%
%
%

\acknowledgments
Construction of the LWA has been supported by the Office of Naval Research under Contract N00014-07-C-0147 and by the AFOSR. Support for operations and continuing development of the LWA1 is provided by the Air Force Research Laboratory and the National Science Foundation under grants AST-1711164, AST-1835400 and AGS-1708855. The authors would like to thank two anonymous referees for 
constructive suggestions. We also thank Jacob Dowell for useful discussions on statistical
comparisons between non-nested models. 
The broadband all-sky image data from LWA-SV used in this article can be found at this site (\url{https://lda10g.alliance.unm.edu/~pasi/MRA_spectra_data/}) The optical data used in this article can be found at Global Meteor Network (\url{https://globalmeteornetwork.org/data/traj_summary_data/monthly/}).

\bibliography{agusample}

\end{document}